# Linear Power System Modeling and Analysis Across Wide Operating Ranges: A Hierarchical Neural State-Space Equation Approach

Weicheng Liu, Mengkai Hu, Di Liu, Songyan Zhang, and Chao Lu, *Senior Member, IEEE*

***Abstract*****—**As modern power systems exhibit increasingly high-dimensional, nonlinear, and uncertain characteristics, the applicability of classical linear state-space methods is severely challenged. Existing paradigms struggle to reconcile the analytical transparency of physics-based models with the continuous nonlinear generalization of AI. To address this, the Hierarchical Neural State-Space Equation (HNSSE) framework is proposed. At the component level, the formulated Neural State-Space Equation (NSSE) extends neural ordinary differential equations (NODEs) to learn continuous dynamic manifolds across varying conditions while strictly preserving local analytical transparency. At the system level, a hierarchical architecture analytically fuses components via network constraints, constructing an interaction-consistent global NODE while circumventing the curse of dimensionality. To ensure robust convergence under noisy measurements, a training strategy synergizing spatiotemporal slicing, physics-informed curriculum learning, and Expectation-Maximization-based refinement is established. Validation on the large-scale Guangdong Power Grid demonstrates the framework's remarkable performance in interpretable state-space reconstruction, high-fidelity trajectory prediction, continuous stability perception, and noise robustness. Comprehensive comparisons substantiate HNSSE's superiority as a unified, interpretable paradigm for complex power system modeling.

## I. Introduction

### *A. Background and motivation*

Linear state-space models have long served as a fundamental tool for power system dynamic analysis and control. Power systems are intrinsically high-dimensional nonlinear dynamical systems; local linearization around a specified operating equilibrium represents a classical modeling paradigm that sacrifices the complete characterization of global nonlinear dynamics in exchange for explicit state-space matrices, clear modal structures, and mature analytical tools. This paradigm has supported key tasks such as small-signal stability analysis, modal identification, damping evaluation, oscillation source localization, and controller design. However, as modern power systems exhibit increasingly high-dimensional, nonlinear, and uncertain characteristics, the applicable operating range of specific linear models is becoming severely restricted [1]. Due to the inevitable deviation between actual nonlinear dynamics and localized approximations, the models may yield biased estimations of critical modal features and stability margins, thereby fundamentally compromising dynamic security assessment and proactive system control. Although mainstream mechanism-based and data-driven methods in the field have proposed enhancements to adapt to these shifts, they still struggle to reconcile the intrinsic trade-off between global generalization and physical transparency [2], [3]. Therefore, modern power system dynamic modeling urgently requires a new paradigm that can continuously characterize complex dynamics evolving across varying operating conditions while strictly preserving the structural transparency, physical interpretability, and the analytical advantages of explicit state-space structures.

### *B. Related work and research gap*

Given the high-dimensional nonlinear nature of power systems, conventional methods typically perform local linearization around specific operating points to approximate complex dynamics via linear state-space models [4]. While this paradigm provides explicit matrices and mature analytical tools [5], finite sets of localized models struggle to accurately cover the broader cross-condition dynamic characteristics of modern power systems [6]. Furthermore, physics-based linear models rely strictly on accurate device parameters and complete control structures; the unavailability or variation of these details inevitably induces modeling errors that compromise stability assessment, controller tuning, and simulation credibility [7]–[9].

To reduce the dependence on complete physical models and accurate parameters, measurement-based identification methods estimate system dynamics using phasor measurement unit (PMU) measurements, disturbance responses, or simulation data. However, existing techniques struggle to fully capture continuous nonlinear evolution across wide operating conditions. Specifically, linear identification and modal decomposition (e.g., Subspace, Prony) produce point-wise linear approximations whose predictive fidelity deteriorates rapidly as operating points drift [10], [11]. Meanwhile, online recursive estimation methods enable localized tracking, yet their continual updates may induce structural forgetfulness, as they adapt to immediate measurements without internalizing global nonlinear laws [12]. Although global dynamics

This work was supported by the Youth Program of the National Natural Science Foundation of China (Grant No. 52407116). (Corresponding author: Chao Lu.)

The authors are with the Department of Electrical Engineering, Tsinghua University, Beijing, China (Email: lwc23@mails.tsinghua.edu.cn, humengkai2003@163.com, kfliudi@mail.tsinghua.edu.cn, zhangsy23@mails.tsinghua.edu.cn, luchao@tsinghua.edu.cn).

discovery methods (e.g., SINDy, Koopman) [13], [14] attempt to expand the representation boundary, they often suffer from the loss of explicit physical state meaning or dimensional explosion. Furthermore, by identifying component dynamics in isolation, these methods systematically ignore network algebraic coupling constraints. Ultimately, the inability to explicitly describe the continuous evolution of state-space matrices severely restricts their cross-condition transferability, fundamentally compromising downstream applications such as online stability assessment, oscillation localization, and proactive dispatching support.

AI-driven approaches further enhance the representation capability for complex nonlinear dynamics. Deep sequence models (e.g., LSTMs, Transformers) can learn cross-condition dynamic patterns from large-scale data and have shown promising performance in dynamic prediction tasks [15], [16]. However, typically taking the form of black-box input-output mappings, these methods struggle to provide physically meaningful state variables, system matrices, and modal structures, rendering them inapplicable for rigorous engineering tasks such as damping evaluation and controller design [17].

Neural ordinary differential equations (NODEs) further introduce continuous-time dynamical structures, making them more suitable for dynamic system modeling than conventional black-box networks and improving structural interpretability to some extent [18]–[21]. However, existing applications of NODEs typically learn black-box vector fields in the latent space of localized devices, lacking the explicit state-space formulation required for downstream applications. Furthermore, by neglecting inherent device-network algebraic couplings, directly extending these methods to large-scale interconnected grids precipitates a curse of dimensionality and fails to guarantee global physical consistency [22]–[24].

In summary, existing methods still leave a fundamental gap. Conventional linearization, physics-based modeling, and measurement-based identification provide good interpretability and compatibility with engineering analysis, but their capability to continuously represent wide-range operating conditions remains limited. Purely data-driven and neural differential equation-based methods offer strong nonlinear representation capability, but they lack globally consistent explicit state-space structures. For large-scale power systems, it is also necessary to avoid the excessive dimensionality, low data efficiency, and loss of device-network coupling consistency caused by monolithic black-box fitting. Therefore, a new modeling framework is needed to continuously represent cross-condition dynamics while preserving the structural interpretability of linear state-space models and supporting hierarchical modeling of large-scale interconnected systems.

### *C. Contributions*

To address these challenges, this paper proposes the Hierarchical Neural State-Space Equation (HNSSE) framework. By mathematically extending standard NODEs, HNSSE deeply integrates physical mechanisms with neural networks to learn continuous nonlinear dynamic manifolds through explicit matrix parameterization. Furthermore, by analytically fusing decentralized component dynamics via network constraints, it circumvents the curse of dimensionality and black-box opacity, ensuring interpretable, high-fidelity reconstruction and robust generalization. The specific contributions are:

**1. To reconcile the intrinsic conflict between the linear state-space structure and nonlinear generalization across operating points, a general Neural State-Space Equation (NSSE) paradigm is formulated.** NSSE is defined as a generalized extension of the NODE structure, thereby obtaining a neural-regularized fundamental form for describing continuous linear system state-space dynamics. This formulation enables the learning of the continuous dynamic manifold of components that evolves with operating conditions. Unlike conventional opaque NODEs, NSSE utilizes neural networks to characterize the evolving state-space dynamics as continuous functions of operating conditions. This design effectively bridges the gap between the opacity of pure AI and the generalization difficulties of traditional identification, achieving a unified expression that combines mechanistic interpretability with data-driven adaptability across wide operating conditions.

**2. To extend the structural consistency of the state-space to the system level based on the dynamic coupling relationships of the power system, and to overcome the curse of dimensionality in large-scale system modeling, a hierarchical mechanistic fusion architecture is designed.** This framework decouples the learning of local component dynamics from global network algebraic constraints. Through mechanism-preserving analytical fusion, decentralized component NSSE models are integrated into an interaction-consistent transparent system-level NODE. This architecture preserves complex global coupling characteristics while reducing modeling complexity from an exponential level to a near-linear relationship with the network scale, thereby ensuring scalability.

**3. To guarantee robust convergence and high-fidelity reconstruction from noisy, multi-scale real-world measurement data, a comprehensive algorithmic training methodology is established.** To accurately capture the complex coupling dynamics of the device-network-system hierarchy, the strategy first synergizes spatiotemporal probabilistic slicing and multi-objective curriculum learning. These techniques orchestrate a dynamic gradient equilibrium across constraints to accelerate stable convergence. Furthermore, to explicitly ensure robustness against measurement noise, an Expectation-Maximization (EM)-based bootstrap refinement strategy is integrated. This mechanism employs iterative optimization to progressively denoise model outputs, ensuring high-precision identification even under heavy interference.

### *D. Paper Organization*

The remainder of this paper is organized as follows. Section II introduces the problem formulation and theoretical preliminaries. Section III details the proposed HNSSE framework. Section IV describes the robust multi-stage learning methodology. Section V presents the numerical validation, and Section VI concludes the paper.

## II. Problem Formulation

### *A. Linear Modeling of Power System Dynamics*

Linearizing the nonlinear differential-algebraic equations (DAEs) of a power system around an operating condition yields the global state-space model:

$$\Delta\dot{\boldsymbol{x}} = \boldsymbol{A}_{\mathrm{sys}}(\boldsymbol{p})\Delta\boldsymbol{x} \tag{1}$$

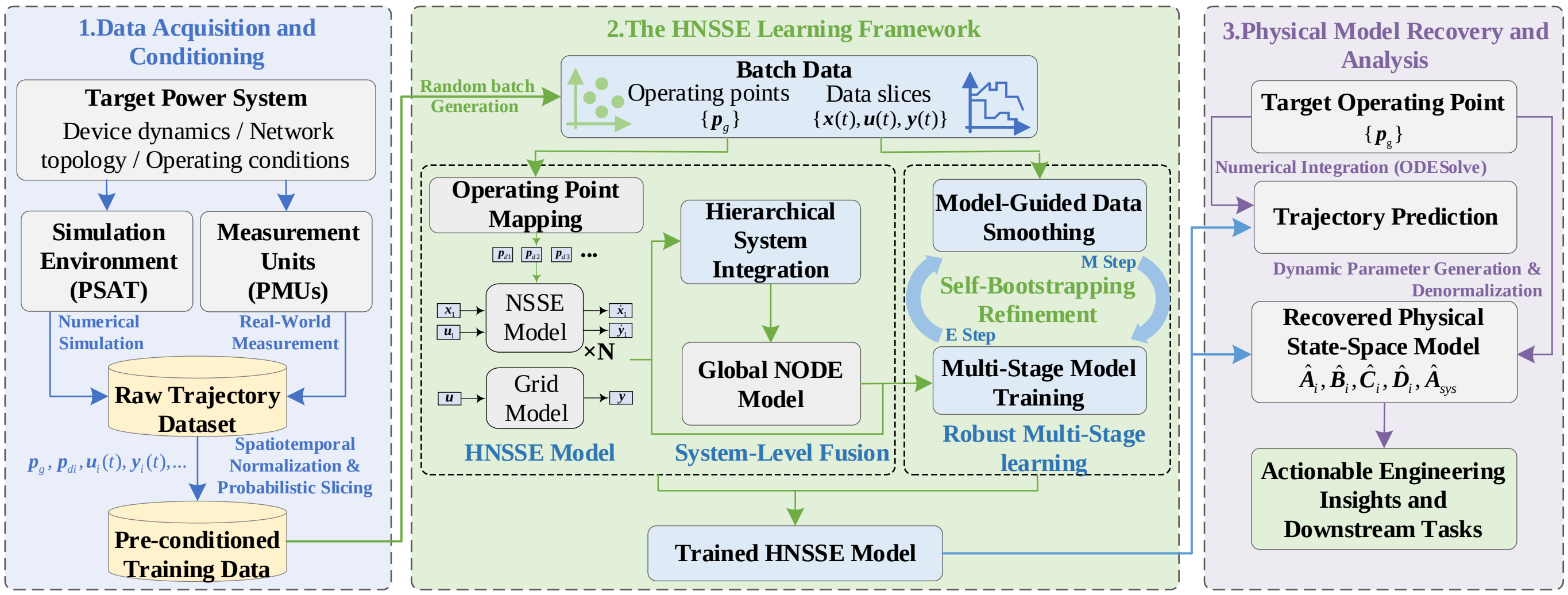

**Fig. 1.** Overall architecture of the proposed HNSSE framework. From left to right, the workflow delineates the process of raw measurement pre-conditioning, robust hierarchical dynamics learning (the theoretical core), and the ultimate recovery of interpretable physical models for downstream engineering tasks.

where $\Delta \boldsymbol{x} = \boldsymbol{x} - \boldsymbol{x}_0$ denotes the state deviation vector. The matrix $\boldsymbol{A}_{sys}(\boldsymbol{p}) \in \mathbb{R}^{n_x \times n_x}$ encapsulates the parameter-dependent modal characteristics. Structurally, $\boldsymbol{A}_{sys}$ is analytically assembled from interconnected components, each operating as a local linear parameter-varying system:

$$\begin{aligned} \Delta \dot{\boldsymbol{x}}_i &= \boldsymbol{A}_i(\boldsymbol{p})\Delta \boldsymbol{x}_i + \boldsymbol{B}_i(\boldsymbol{p})\Delta \boldsymbol{u}_i \\ \Delta \boldsymbol{y}_i &= \boldsymbol{C}_i(\boldsymbol{p})\Delta \boldsymbol{x}_i + \boldsymbol{D}_i(\boldsymbol{p})\Delta \boldsymbol{u}_i \end{aligned} \tag{2}$$

where $\Delta \boldsymbol{x}_i$, $\Delta \boldsymbol{u}_i$, and $\Delta \boldsymbol{y}_i$ represent deviations in the $i$-th component's internal states, inputs, and outputs, respectively, coupled via network algebraic constraints.

**Problem Statement:** In the context of data-driven modeling, the problem is reformulated as follows:

Given a dataset of system-wide measurement trajectories collected across diverse operating conditions $\mathcal{D} = \{(\boldsymbol{x}^{(k)}, \boldsymbol{u}^{(k)}, \boldsymbol{y}^{(k)}, \boldsymbol{p}^{(k)})\}$, the goal is to learn a parameterized framework that can: **1) identify** the device-level linear dynamics as continuous functions of $\boldsymbol{p}$; **2) synthesize** the global system matrix $\boldsymbol{A}_{sys}(\boldsymbol{p})$ by fusing local models with known network topology; **3) preserve** the physical interpretability of state variables and the algebraic structure of the power network.

To address this, we propose the HNSSE framework, outlined in Fig. 1 and detailed in Sections III and IV.

*B. Preliminaries: Data-Driven Dynamic Modeling with Neural Ordinary Differential Equations*

The modeling task formulated above necessitates a learning paradigm capable of capturing continuous-time dynamics from sampled time-series data. NODEs offer a rigorous mathematical framework for this purpose [18]. A standard NODE approximates the unknown vector field $\boldsymbol{f}$ governing state evolution $\dot{\boldsymbol{x}}(t)$ using a neural network $\boldsymbol{f}_{\theta}$:

$$\dot{\boldsymbol{x}}(t) = \boldsymbol{f}_{\theta}(t, \boldsymbol{x}) \tag{3}$$

By integrating an explicit ODE solver, the state trajectory is predicted from an initial state $\boldsymbol{x}(t_0)$:

$$\boldsymbol{x}(t_k) = \text{ODESolve}(\boldsymbol{x}(t_0), \boldsymbol{f}_{\theta}, t_0, t_k) \tag{4}$$

The parameters $\boldsymbol{\theta}$ are optimized by backpropagating gradients through the solver—using direct backpropagation or the adjoint sensitivity method — to minimize trajectory prediction error. While standard NODEs model autonomous systems as black boxes, Section III extends this foundation to incorporate exogenous inputs, hierarchical topologies, and explicit state-space structural constraints essential for interpretable power system modeling.

## III. The Hierarchical Neural State-Space Equation Framework

*A. The Neural State-Space Equation*

The NSSE constitutes the atomic unit of the proposed framework, whose architecture is depicted in Fig. 2. Conceptually, it is defined as a structural regularization and generalized extension of the standard NODE paradigm. While standard NODEs approximate an arbitrary black-box vector field $\dot{\boldsymbol{x}} = f_{\theta}(t, \boldsymbol{x})$, the NSSE imposes a hard mathematical constraint to enforce the linear state-space form.

For a generic dynamic component $i$, the NSSE defines its dynamics as an affine transformation field conditioned on the local operating point $\boldsymbol{p}_{d,i}$. The general continuous-time dynamics are formulated as:

$$\begin{aligned} \Delta \dot{\boldsymbol{x}}_i &= \boldsymbol{f}_{\theta}(\Delta \boldsymbol{x}_i, \boldsymbol{p}_{d,i}) + \boldsymbol{g}_{\phi}(\Delta \boldsymbol{u}_i, \boldsymbol{p}_{d,i}) \\ \Delta \boldsymbol{y}_i &= \boldsymbol{h}_{\psi}(\Delta \boldsymbol{x}_i, \boldsymbol{p}_{d,i}) + \boldsymbol{k}_{\omega}(\Delta \boldsymbol{u}_i, \boldsymbol{p}_{d,i}) \end{aligned} \tag{5}$$

To strictly enforce the interpretability required by linear control theory, we restrict the neural functions $\boldsymbol{f}, \boldsymbol{g}, \boldsymbol{h}, \boldsymbol{k}$ to be affine transformations. Consequently, the component dynamics collapse into the standard state-space representation, generalized to an affine form to accommodate the non-zero equilibrium manifolds inherent in practical operation:

$$\begin{aligned} \dot{\boldsymbol{x}}_i &= \boldsymbol{A}_i(\boldsymbol{p}_{d,i})\Delta \boldsymbol{x}_i + \boldsymbol{B}_i(\boldsymbol{p}_{d,i})\Delta \boldsymbol{u}_i \\ &= \boldsymbol{A}_i(\boldsymbol{p}_{d,i})\boldsymbol{x}_i + \boldsymbol{B}_i(\boldsymbol{p}_{d,i})\boldsymbol{u}_i + \boldsymbol{b}_{x,i}(\boldsymbol{p}_{d,i}) \\ \boldsymbol{y}_i &= \boldsymbol{y}_{i,0} + \boldsymbol{C}_i(\boldsymbol{p}_{d,i})\Delta \boldsymbol{x}_i + \boldsymbol{D}_i(\boldsymbol{p}_{d,i})\Delta \boldsymbol{u}_i \\ &= \boldsymbol{C}_i(\boldsymbol{p}_{d,i})\boldsymbol{x}_i + \boldsymbol{D}_i(\boldsymbol{p}_{d,i})\boldsymbol{u}_i + \boldsymbol{b}_{y,i}(\boldsymbol{p}_{d,i}) \end{aligned} \tag{6}$$

This formulation establishes a direct isomorphism between the neural weights and the physical system matrices. Specifically, the system matrices $\mathcal{S}_i = \{\boldsymbol{A}_i, \boldsymbol{B}_i, \boldsymbol{C}_i, \boldsymbol{D}_i\}$ correspond directly to the weight matrices of the affine layers, while the bias vectors $\mathcal{B}_i = \{\boldsymbol{b}_{x,i}, \boldsymbol{b}_{y,i}\}$ capture the shift in equilibrium points.

The adaptability of the NSSE is realized through a dynamic parameterization mechanism. The system matrices $\mathcal{S}_i$ and biases $\mathcal{B}_i$ are not static parameters; they are continuous functions of the operating condition $\boldsymbol{p}_{d,i}$ . We model this dependency using Parameter-Generating Networks (PGNs), denoted as $\Phi$ :

$$\begin{aligned}\operatorname{vec}(\mathcal{S}_i) &= \Phi_{\mathcal{S}}\left(\boldsymbol{p}_{d,i};\theta_{\mathcal{S}}\right)\\ \mathcal{B}_i &= \Phi_{\mathcal{B}}\left(\boldsymbol{p}_{d,i};\theta_{\mathcal{B}}\right)\end{aligned} \tag{7}$$

where $\operatorname{vec}(\cdot)$ denotes vectorization, and $\theta$ represents the learnable meta-parameters. This architecture separates the linear dynamics manifold (6) from the non-linear parameter manifold (7), allowing the model to capture complex nonlinearities across wide operating ranges while remaining locally linear and analytically tractable.

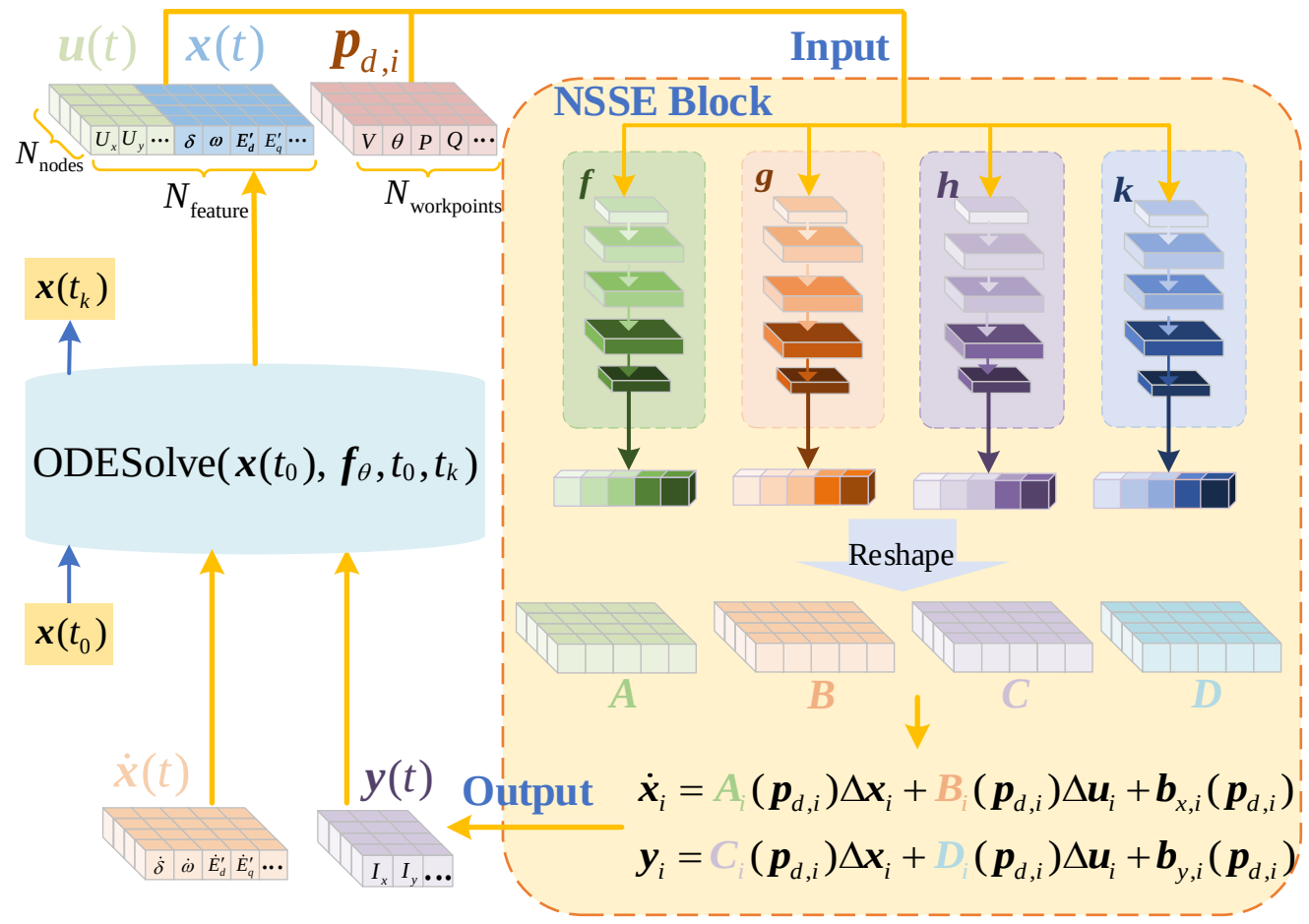

**Fig. 2.** Architecture of the NSSE block.

To ensure the framework operates from a unified system-level input without requiring auxiliary knowledge of individual device-level conditions, we implement a global-to-local mapping strategy. Local operating point $\boldsymbol{p}_{d,i}$ can be inferred from the global system operating point $\boldsymbol{p}_g$ via a learnable mapping network, $\mathcal{N}_{map,i}$ :

$$\boldsymbol{p}_{d,i} = \mathcal{N}_{map,i}\left(\boldsymbol{p}_g;\theta_{map,i}\right) \tag{8}$$

This network learns to approximate the relationship between the global operating points and local conditions for each device, ensuring the HNSSE framework functions as a cohesive, end-to-end model driven solely by global system observables.

The complete forward process integrates these generated components. Given an initial state $\Delta\boldsymbol{x}_i(t_0)$ and an input trajectory $\Delta\boldsymbol{u}_i(t)$ , the state evolution is predicted by numerically integrating the non-autonomous ODE:

$$\Delta\boldsymbol{x}_i(t_k) = \Delta\boldsymbol{x}_i(t_0) + \int_{t_0}^{t_k}\left(\boldsymbol{A}_i(\boldsymbol{p}_{d,i})\Delta\boldsymbol{x}_i(\tau) + \boldsymbol{B}_i(\boldsymbol{p}_{d,i})\Delta\boldsymbol{u}_i(\tau) + \boldsymbol{b}_{x,i}\left(\boldsymbol{p}_{d,i}\right)\right)d\tau \tag{9}$$

The NSSE paradigm thus inherits the principle of learning continuous dynamics from NODEs but critically extends this foundation with an explicit, physically interpretable input-output state-space structure. This endows the NSSE with a dual capability: it operates as a fully interpretable, standalone model whose consistent state-space structure supports component-level stability analysis and decentralized control design, while simultaneously serving as a standardized algebraic building block for the system-level mechanism fusion described in Section III-B.

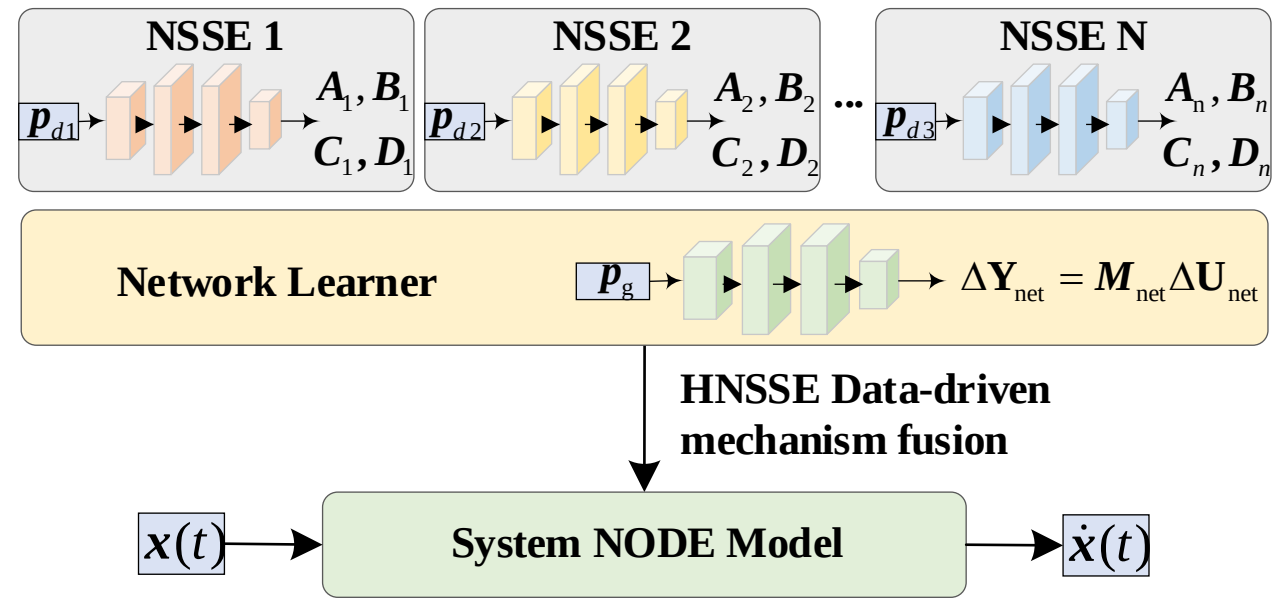

**Fig. 3.** The fusion architecture of the system-level NODE.

## B. Hierarchical System Integration and the Global Model

The HNSSE framework achieves scalability and system-wide coherence through a compositional modeling approach that facilitates a data-driven mechanism fusion. This is accomplished by separately modeling the dynamic devices and their network interconnections, and then analytically integrating these representations to form a global system model.

The interconnections between components are described by the power network's algebraic equations. When linearized, these equations yield a linear relationship between the stacked vectors of all component port inputs, $\Delta\boldsymbol{U}_{\text{net}}$ , and outputs, $\Delta\boldsymbol{Y}_{\text{net}}$ :

$$\Delta\boldsymbol{Y}_{\text{net}} = \boldsymbol{M}_{\text{net}}\Delta\boldsymbol{U}_{\text{net}} + \boldsymbol{b}_{net} \tag{10}$$

The physical meaning of the network matrix $\boldsymbol{M}_{\text{net}}$ depends on the choice of interface variables. For instance, if the real and imaginary parts of port voltages and currents are selected, $\boldsymbol{M}_{\text{net}}$ represents the system's nodal admittance matrix, which is constant and independent of the operating point. However, to better match practical measurements, this work utilizes voltage magnitude and angle deviations ( $\Delta\boldsymbol{V}$ , $\Delta\boldsymbol{\theta}$ ) as inputs, and active and reactive power deviations ( $\Delta\boldsymbol{P}$ , $\Delta\boldsymbol{Q}$ ) as outputs. Consequently, $\boldsymbol{M}_{\text{net}}$ corresponds to the power flow Jacobian matrix, which inherently varies nonlinearly with the operating point. To preserve generalization capabilities, we employ a priori physical relationship embedding to map the learning parameters to the operating-point-independent equivalent real-valued admittance matrix $\boldsymbol{Y}$ . This matrix defines the fundamental linear relationship when the inputs and outputs are chosen as the rectangular coordinates (real and imaginary parts) of voltages $\boldsymbol{U}_{xy} = [\boldsymbol{V}_x^T, \boldsymbol{V}_y^T]^T$ and currents $\boldsymbol{I}_{xy} = [\boldsymbol{I}_x^T, \boldsymbol{I}_y^T]^T$ :

$$\boldsymbol{I}_{xy} = \boldsymbol{Y}\boldsymbol{U}_{xy}, \quad \text{where } \boldsymbol{Y} = \begin{bmatrix}\boldsymbol{G} & -\boldsymbol{B}\\ \boldsymbol{B} & \boldsymbol{G}\end{bmatrix} \tag{11}$$

During the training and inference phases, the corresponding network matrix $\boldsymbol{M}_{\text{net}}$ is then mechanistically derived based on the specific operating point:

$$\boldsymbol{M}_{\text{net}} = \begin{bmatrix}\dfrac{\partial\boldsymbol{P}}{\partial\boldsymbol{\theta}} & \dfrac{\partial\boldsymbol{P}}{\partial\boldsymbol{V}}\\[2mm] \dfrac{\partial\boldsymbol{Q}}{\partial\boldsymbol{\theta}} & \dfrac{\partial\boldsymbol{Q}}{\partial\boldsymbol{V}}\end{bmatrix}, \text{e.g., } \frac{\partial P_i}{\partial\theta_j} = V_iV_j(G_{ij}\sin\theta_{ij} - B_{ij}\cos\theta_{ij}) \tag{12}$$

Consequently, the learned network relationships remain consistent across varying operating points, while the dynamic evolution of $\boldsymbol{M}_{\text{net}}$ is governed by these mechanistic equations. This formulation fully ensures generalization while satisfying end-to-end differentiability, allowing system-level errors to be directly backpropagated to the underlying parameterized admittance matrix.

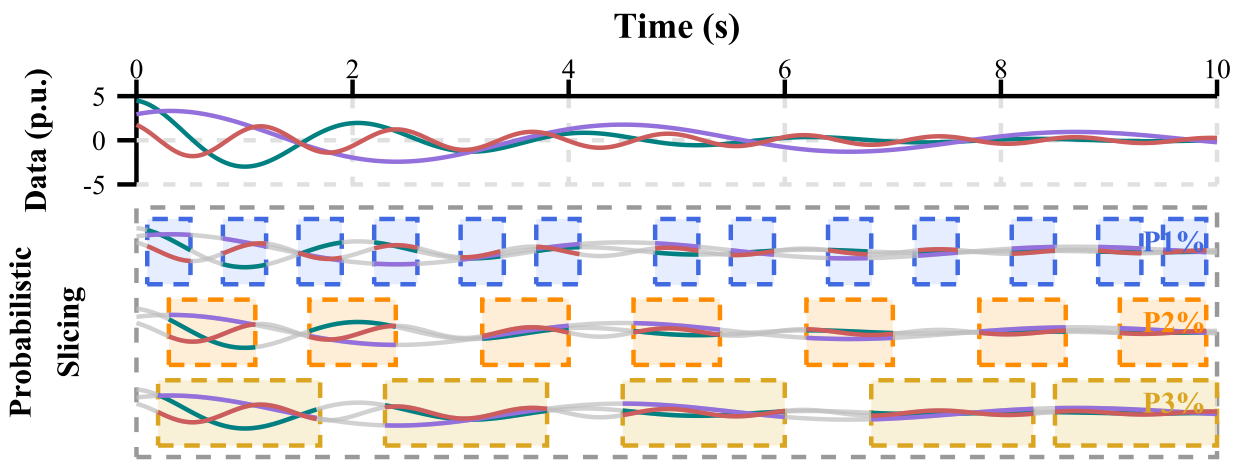

**Fig. 4.** Illustration of the probabilistic slicing strategy for capturing multi-scale dynamics from a decaying waveform.

With data-driven models for both the individual components (the set of NSSEs) and the network (the matrix $\boldsymbol{M}_{\text{net}}$), the global system state matrix $\boldsymbol{A}_{\text{sys}}$ can be analytically constructed. First, the matrices from all component NSSE models are aggregated into block-diagonal system matrices:

$$\begin{aligned} &\boldsymbol{A} = \text{diag}(\boldsymbol{A}_1, \boldsymbol{A}_2, \ldots, \boldsymbol{A}_N) \quad \boldsymbol{B} = \text{diag}(\boldsymbol{B}_1, \boldsymbol{B}_2, \ldots, \boldsymbol{B}_N) \\ &\boldsymbol{C} = \text{diag}(\boldsymbol{C}_1, \boldsymbol{C}_2, \ldots, \boldsymbol{C}_N) \quad \boldsymbol{D} = \text{diag}(\boldsymbol{D}_1, \boldsymbol{D}_2, \ldots, \boldsymbol{D}_N) \\ &\boldsymbol{b}_{state} = [\boldsymbol{b}_{x,1}^T, \ldots, \boldsymbol{b}_{x,N}^T]^T \quad \boldsymbol{b}_{output} = [\boldsymbol{b}_{y,1}^T, \ldots, \boldsymbol{b}_{y,N}^T]^T \end{aligned} \tag{13}$$

These are then combined with the network matrix through the standard algebraic elimination of the interface variables, yielding the final, interpretable global state matrix:

$$\begin{aligned} \boldsymbol{A}_{\text{sys}} &= \boldsymbol{A} + \boldsymbol{B}(\boldsymbol{M}_{\text{net}} - \boldsymbol{D})^{-1}\boldsymbol{C} \\ \boldsymbol{b}_{sys} &= \boldsymbol{b}_{state} + \boldsymbol{B}(\boldsymbol{M}_{net} - \boldsymbol{D})^{-1}(\boldsymbol{b}_{output} - \boldsymbol{b}_{net}) \end{aligned} \tag{14}$$

The HNSSE framework achieves a unique synthesis of mechanistic principles and deep learning capabilities, as illustrated in Fig. 3. The framework's adherence to the ODE-based state-space structure is the key to ensuring a physically interpretable mechanism fusion through the analytical assembly process. Simultaneously, the parameterization of all component matrices via the neural network form preserves end-to-end differentiability, ensuring seamless gradient propagation from a system-level objective back to each local model for true co-optimization. This hybrid design—where a classical ODE structure is parameterized by deep neural networks—is the critical feature that allows the complete system model to be treated as a **large, interpretable system-level NODE**. Its dynamic evolution is governed by the learned global vector field $\Delta\dot{\boldsymbol{x}}_{sys} = \boldsymbol{f}_{sys}(\Delta\boldsymbol{x}_{sys}, \boldsymbol{p}_g) = \boldsymbol{A}_{sys}(\boldsymbol{p}_g)\Delta\boldsymbol{x}_{sys} + \boldsymbol{b}_{sys}(\boldsymbol{p}_g)$. The state trajectory can thus be predicted through standard autonomous numerical integration:

$$\Delta\boldsymbol{x}_{\text{sys}}(t_k) = \text{ODESolve}(\Delta\boldsymbol{x}_{\text{sys}}(t_0), \boldsymbol{f}_{\text{sys}}, t_0, t_k) \tag{15}$$

## IV. Robust Multi-Stage Learning Framework and Physical Recovery

### A. Dataset Formulation and Spatiotemporal Preprocessing

The HNSSE framework is trained on a collection of small-signal time-series trajectories. The raw dataset $\mathcal{D}_{raw}$ is defined as a set of $N$ samples, where each sample corresponds to the system's dynamic response under a specific global operating condition $\boldsymbol{p}_{g,n}$:

$$\mathcal{D}_{raw} = \{(\boldsymbol{p}_{g,n}, \boldsymbol{u}_n(t), \boldsymbol{y}_n(t), \boldsymbol{x}_n(t))\}_{n=1}^{N} \tag{16}$$

Here, $\boldsymbol{u}_n, \boldsymbol{y}_n, \boldsymbol{x}_n$ represent the raw trajectories with heterogeneous physical units. To ensure numerical stability, we apply a Spatiotemporal Normalization strategy. For any variable type $v \in \{x, u, y\}$, let $\boldsymbol{\mu}_{v,n}$ denote the sample-specific reference offset (typically derived from the steady-state measurement) used to center the trajectory, and $\boldsymbol{S}_v = \text{diag}(\sigma_v)$ denote the global diagonal standard deviation matrix. The normalized dataset $\mathcal{D}$ is obtained via the transformation:

$$\tilde{\boldsymbol{v}}_n(t) = \boldsymbol{S}_v^{-1}(\boldsymbol{v}_n(t) - \boldsymbol{\mu}_{v,n}) \tag{17}$$

The transformation parameters $\{\boldsymbol{S}_v, \boldsymbol{\mu}_{v,n}\}$ are preserved to facilitate the analytical physical model recovery detailed in Section IV-D.

Subsequently, to capture multi-scale dynamics and mitigate optimization bias towards initial transients, we employ a Probabilistic Slicing strategy. As illustrated in Fig. 4, this strategy augments the dataset by sampling sub-trajectories of stochastic lengths from $\mathcal{D}$, ensuring the model is exposed to a balanced spectrum of transient and steady-state modes.

### B. Physics-Informed Multi-Objective Curriculum Learning

Training the hierarchical architecture constitutes a complex optimization problem that requires balancing three distinct objectives: satisfying network algebraic constraints, fitting local component dynamics, and reproducing system-level emergent behavior. To facilitate the formulation of the multi-objective loss, we explicitly define the four key quantities predicted by the HNSSE framework based on the normalized inputs: 1. $\Delta\hat{\boldsymbol{x}}_i(t)$ : Predicted local state trajectory of component $i$; 2. $\Delta\hat{\boldsymbol{y}}_i(t)$: Predicted local output trajectory; 3. $\Delta\hat{\boldsymbol{x}}_{sys}(t)$ : Predicted global system state trajectory; 4. $\Delta\hat{\boldsymbol{y}}_{net}(t)$ : Predicted global network output trajectory. To navigate the complex optimization landscape and avoid gradient conflicts, we formulate a Physics-Informed Multi-Objective Loss and employ a Two-Stage Curriculum Learning strategy.

**1) The Multi-Objective Fusion Loss:** The total loss function $\mathcal{L}_{total}$ is defined as a dynamic weighted sum of three components:

$$\mathcal{L}_{total} = \alpha(t)\mathcal{L}_{net} + \beta(t)\mathcal{L}_{local} + \gamma(t)\mathcal{L}_{global} \tag{18}$$

where $\alpha, \beta, \gamma$ are time-dependent weighting coefficients. The components are defined as follows:

**Network Constraint Loss ($\mathcal{L}_{net}$):** Enforces the network algebraic coupling constraints at the interconnection ports. It penalizes the residual between the network output $\Delta\tilde{\boldsymbol{y}}_{net}$ and the physical response computed via the network matrix $\boldsymbol{M}_{net}$:

$$\mathcal{L}_{net} = \frac{1}{N}\sum_{n=1}^{N} \| \Delta\boldsymbol{y}_{net,n} - \Delta\hat{\boldsymbol{y}}_{net,n} \|_2^2 \tag{19}$$

**Local Fidelity Loss ($\mathcal{L}_{local}$):** Ensures that each NSSE accurately captures the input-state-output behavior of its specific component. It aggregates the errors of state trajectories and interface outputs across all $M$ devices:

$$\mathcal{L}_{local} = \frac{1}{NM}\sum_{n=1}^{N}\sum_{i=1}^{M}(\| \Delta\tilde{\boldsymbol{x}}_{i,n} - \Delta\hat{\boldsymbol{x}}_{i,n} \|_2^2 + \lambda \| \Delta\tilde{\boldsymbol{y}}_{i,n} - \Delta\hat{\boldsymbol{y}}_{i,n} \|_2^2) \tag{20}$$

**Global Coherency Loss ($\mathcal{L}_{global}$):** Enforces the consistency of the analytically fused system-level model. It minimizes the divergence between the observed global trajectory and the integration result of the synthesized global vector field:

$$\mathcal{L}_{global} = \frac{1}{N}\sum_{n=1}^{N} \| \Delta\boldsymbol{x}_{sys,n} - \Delta\hat{\boldsymbol{x}}_{sys,n} \|_2^2 \tag{21}$$

**2) Two-Stage Curriculum Strategy:** Directly optimizing all objectives from random initialization often leads to convergence issues due to the stiffness difference between local and global dynamics. Therefore, we design a progressive curriculum:

**Stage I. Optional Pre-structuring and Decoupled Learning ( $\gamma \approx 0$ ):** In the initial phase, the focus is strictly on learning the local vector fields and satisfying algebraic constraints. The HNSSE architecture supports optional structural initialization, allowing the model to be warm-started based on available prior knowledge: **1. Topology-Informed:** If the network structure is known, the equivalent admittance matrix $\boldsymbol{Y}$ is fixed to the theoretical admittance matrix to provide a hard physical constraint; otherwise, it is treated as a learnable parameter; **2. Parameter-Informed:** For components with known typical parameters (e.g., standard generator models), the weights of the NSSE can be initialized to approximate these nominal dynamics, accelerating convergence; **3. Structure-Informed:** Known kinematic relationships (e.g., rotor angle derivative equals speed) can be frozen within the NSSE matrix structure. This stage essentially performs parallel identification of decoupled components, rapidly establishing a coarse-grained model representation.

**Stage II. Global-Local Fusion Learning: ( $\gamma \to \gamma_{max}$ ):** Once the local models stabilize, we transition to a holistic training phase by gradually increasing the weight $\gamma$ while retaining significant weights for $\mathcal{L}_{local}$ and $\mathcal{L}_{net}$ . This step activates the Mechanism Fusion layer, enabling the backpropagation of system-level gradients to adjust local parameters. By simultaneously optimizing local fidelity and global coherency, this stage ensures that the individual components are not only locally accurate but also mathematically compatible to reproduce the complex, coupled oscillatory modes of the interconnected system. This curriculum-based approach stabilizes the training process, accelerates convergence, and leads to a more accurate and physically consistent final model.

### *C. EM-Based Self-Bootstrapping Refinement*

In real-world scenarios, PMU measurements are inevitably corrupted by sensor noise. The proposed HNSSE framework inherently possesses a significant degree of noise resilience: the probabilistic slicing strategy enables the model to capture long-range dependencies, while the numerical integration process naturally acts as a low-pass filter, smoothing out high-frequency perturbations. Furthermore, the global-local fusion mechanism maximizes the utilization of redundant information across the network. However, to further elevate the modeling fidelity and extract the cleanest possible physical parameters from corrupted data, we introduce a Self-Bootstrapping Refinement strategy rooted in the EM paradigm.

**1) Probabilistic State-Space Formulation:** We treat the true, noise-free state trajectories $\boldsymbol{X} = \{\boldsymbol{x}_0, \dots, \boldsymbol{x}_T\}$ as latent variables. Although the HNSSE learns a continuous-time vector field, the measurement data consists of discrete samples. To perform denoising on these discrete measurements based on the learned continuous model, the dynamics are discretized at the sampling instants, yielding the following recursive state-space formulation:

$$\begin{aligned} \boldsymbol{x}_{t+1} &= \boldsymbol{\Phi}\left(\boldsymbol{p}_g;\theta\right)\boldsymbol{x}_t + \boldsymbol{w}_t,\, \boldsymbol{w}_t \sim N\left(0,\boldsymbol{Q}\right) \\ \boldsymbol{z}_t &= \boldsymbol{H}\left(\boldsymbol{p}_g;\theta\right)\boldsymbol{x}_t + \boldsymbol{v}_t,\, \boldsymbol{v}_t \sim N\left(0,\boldsymbol{R}\right) \end{aligned} \tag{22}$$

where $\boldsymbol{z}_t$ denotes the noisy measurement vector. The state transition matrix $\boldsymbol{\Phi}$ is derived from the learned continuous global system matrix $\boldsymbol{A}_{sys}$ in (14) according to the numerical integration scheme employed in the HNSSE forward pass. Taking the Implicit Euler method with step size $\Delta t$ as an example, the transition is defined as $\boldsymbol{\Phi} = (\boldsymbol{I} - \boldsymbol{A}_{sys}(\boldsymbol{p}_g)\Delta t)^{-1}$ . The measurement matrix $\boldsymbol{H}$ and noise covariances $\boldsymbol{Q}, \boldsymbol{R}$ are estimated iteratively.

**2) The Iterative Refinement Loop:** The algorithm alternates between estimating the optimal state trajectories given the current model (E-Step) and optimizing the model parameters given the estimated trajectories (M-Step).

**E-Step (Model-Guided Smoothing):** Given the current parameters $\theta^{(k)}$, we compute the posterior distribution of the latent states $p(\boldsymbol{X} \mid \boldsymbol{Z}, \theta^{(k)})$ . This is achieved using the Rauch-Tung-Striebel (RTS) Smoother, a forward-backward recursive algorithm: **1. Forward Pass (Filtering)***:* Computes the predicted state mean $\hat{\boldsymbol{x}}_{t|t}$ and covariance $\boldsymbol{P}_{t|t}$ incorporating measurements up to time $t$ ; **2. Backward Pass (Smoothing):** Refines these estimates using future measurements to obtain the smoothed state $\hat{\boldsymbol{x}}_{t|T}^{smooth}$ . By exploiting the temporal consistency of the learned physical model, this step effectively filters out Gaussian noise that does not conform to the system dynamics.

**M-Step (Parameter Optimization):** We maximize the expected log-likelihood of the joint distribution. In the context of HNSSE training, this translates to fine-tuning the model using the Stage II Global-Local Fusion curriculum described in Section IV-B. Specifically, the neural network parameters $\theta$ are updated to minimize the total loss $\mathcal{L}_{total}$ (18), but with the noisy raw trajectories in the training set replaced by the smoothed estimates $\hat{\boldsymbol{x}}_{t|T}^{smooth}$ :

$$\theta^{(k+1)} = \arg\min_{\theta} \mathcal{L}_{total}\left(\mathcal{D}_{target}\left(\hat{\boldsymbol{x}}_{t|T}^{smooth}\right);\theta\right) \tag{23}$$

Simultaneously, the noise covariances $\boldsymbol{Q}$ and $\boldsymbol{R}$ are updated based on the residuals of the smoothed estimates, adaptively tuning the filter's bandwidth to match the model's evolving fidelity.

This iterative procedure creates a virtuous cycle: a more accurate physical model enables more precise noise rejection, which in turn provides higher-quality supervision signals for further model refinement.

### *D. Physical Model Recovery and Interpretability*

Upon convergence of the training curriculum, the HNSSE framework yields a predictive model optimized within the dimensionless latent space defined in Section IV-A. To derive actionable engineering insights, the learned normalized parameters must be analytically mapped back to the physical domain.

Let $\tilde{\boldsymbol{S}}_i = \{\tilde{\boldsymbol{A}}_i, \tilde{\boldsymbol{B}}_i, \tilde{\boldsymbol{C}}_i, \tilde{\boldsymbol{D}}_i\}$ denote the system matrices inferred by the NSSE for component $i$ in the normalized space. Leveraging the scaling matrices $\boldsymbol{S}_v$ (where $v \in \{x,u,y\}$ ) preserved from the preprocessing stage, the recovery of the physical state-space matrices $\mathcal{S}_i^{phy}$ is governed by the following linear similarity transformations:

$$\begin{aligned} \boldsymbol{A}_i^{phy} &= \boldsymbol{S}_x \boldsymbol{A}_i \boldsymbol{S}_x^{-1}, \quad \boldsymbol{B}_i^{phy} = \boldsymbol{S}_x \boldsymbol{B}_i \boldsymbol{S}_u^{-1} \\ \boldsymbol{C}_i^{phy} &= \boldsymbol{S}_y \boldsymbol{C}_i \boldsymbol{S}_x^{-1}, \quad \boldsymbol{D}_i^{phy} = \boldsymbol{S}_y \boldsymbol{D}_i \boldsymbol{S}_u^{-1} \end{aligned} \tag{24}$$

Applying the same transformation logic to the fused system-level model, we obtain the global physical system matrix:

$$\boldsymbol{A}_{sys}^{phy}(\boldsymbol{p}_g) = \boldsymbol{S}_x \boldsymbol{A}_{sys}^{norm}(\boldsymbol{p}_g) \boldsymbol{S}_x^{-1} \tag{25}$$

The recovered matrix $\boldsymbol{A}_{sys}^{phy}$ is a data-driven, yet fully interpretable, estimate of the true system state matrix at the specified operating point $\boldsymbol{p}_g$ .

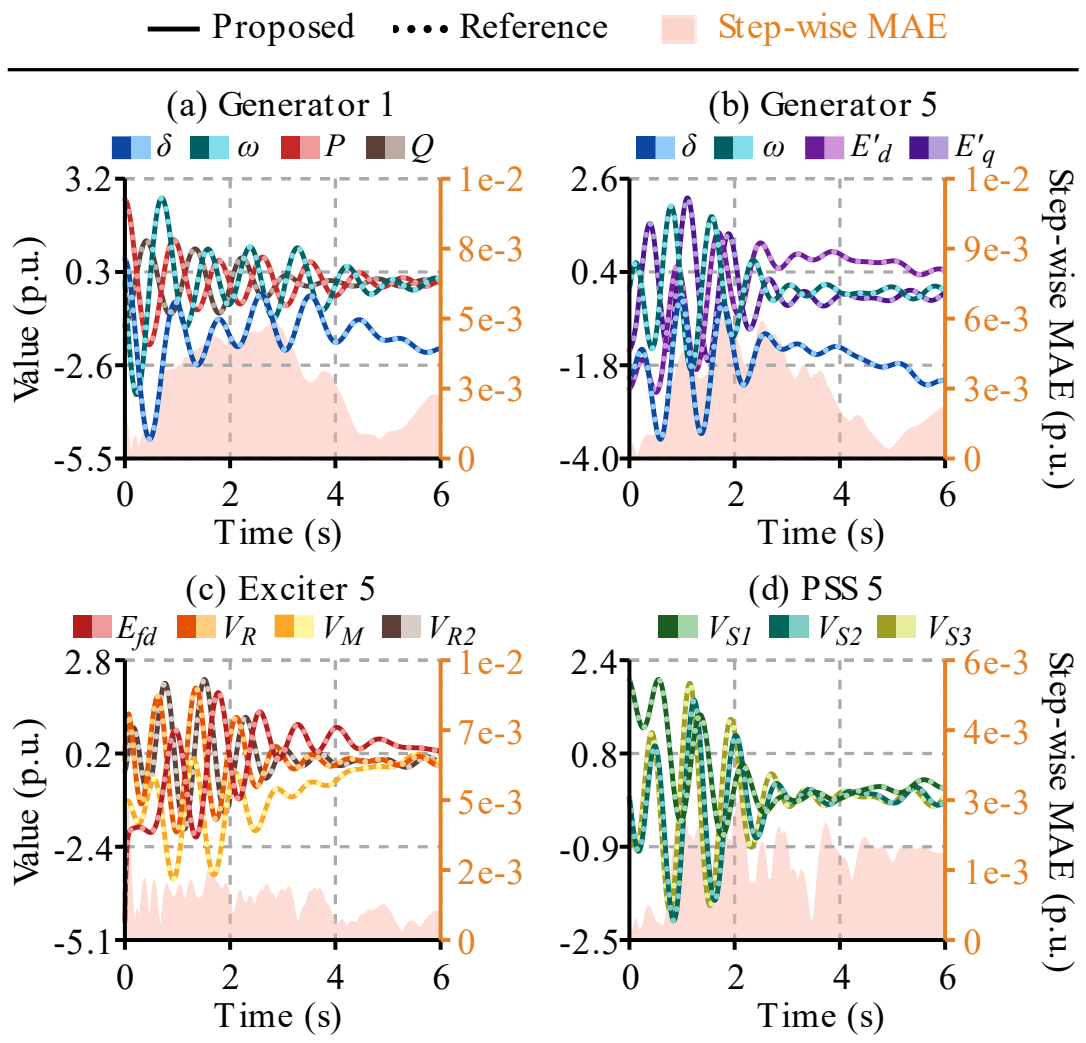

**Fig. 5.** Global trajectory prediction performance of the hierarchically integrated system model.

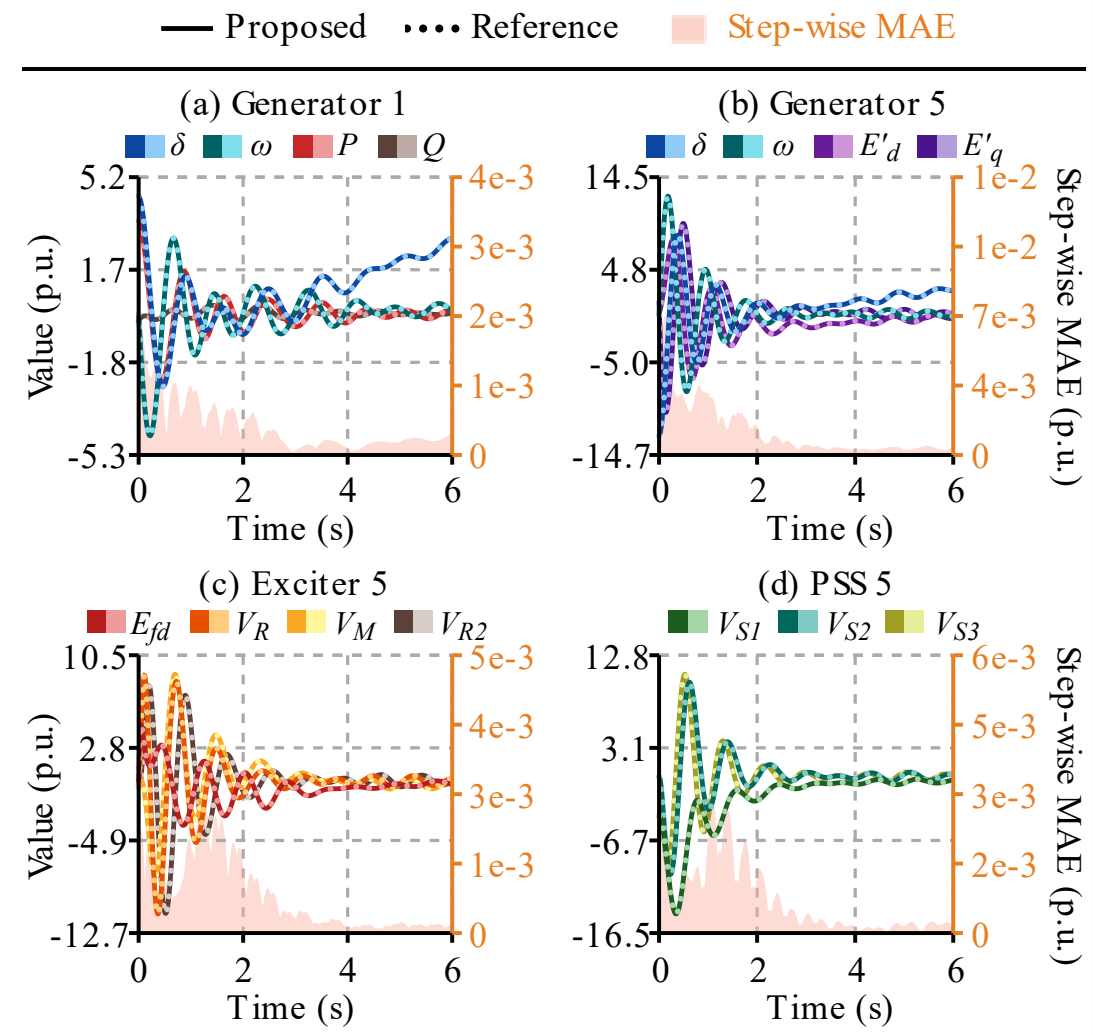

**Fig. 6.** Trajectory prediction of localized states for individual system components under perturbations.

This explicit restoration of the canonical state-space form bridges the gap between black-box deep learning and rigorous engineering requirements. It provides a transparent, standardized mathematical foundation that is inherently compatible with classical control theory, thereby enabling a broad spectrum of downstream applications ranging from small-signal stability assessment to model-based controller design.

## V. Numerical Validation

### *A. Experimental Setup*

To systematically evaluate the proposed HNSSE framework's modeling accuracy, explicit physical reconstruction, and robustness, comprehensive validations are conducted on an 83-bus Guangdong Power Grid model. Ground-truth dynamic datasets are generated using the MATLAB-based Power System Analysis Toolbox (PSAT). To capture the system's nonlinear evolutionary manifold, 200 dynamic simulations (10 s duration, 0.01 s sampling rate mimicking standard PMU reporting) are generated by stochastically varying renewable and load active/reactive power within a ±40% range around the nominal operating point, subject to random state perturbations. The dataset is partitioned into an 8:2 training-testing split to rigorously evaluate wide-condition generalization. Generator 5 serves as the reference machine for relative phase angle variations.

The framework is implemented in PyTorch and executed on a high-performance server (utilizing 10 cores of an Intel Xeon 8470 CPU, an NVIDIA A800 GPU, and 1 TB RAM). The architecture strictly adheres to Section III. Hypernetworks parameterizing the operating-point-dependent matrices consist of 3-hidden-layer MLPs (hidden dimension 32, SiLU activation). Training employs a hybrid optimization strategy [25]: the semi-second-order Muon optimizer for 2D weight tensors to enhance spectral stability, and AdamW for 1D vectors (biases/gains), both utilizing a 0.001 weight decay and dynamic learning rate scheduling.

### *B. Modeling Validation and Structural Interpretability of the HNSSE Framework*

#### 1) Validation of Dynamic Modeling Accuracy Across Shifting Equilibria

The training of the HNSSE framework is principally driven by the objective of minimizing the discrepancy between predicted and ground-truth time-series trajectories. At the foundational level, the proposed model accurately reproduces component-specific dynamic signatures. As illustrated in Fig. 6, the decentralized component NSSEs accurately predict the localized states of various devices under perturbations. As shown in Fig. 6(a) and Fig. 6(b), the model precisely captures the low-frequency electromechanical swing dynamics of the synchronous machines. Simultaneously, for the controllers shown in Fig. 6(c) and (d), it effectively tracks the fast-timescale, control-dominated transients introduced by the Exciter and the Power System Stabilizer (PSS). This demonstrates the framework's capability to accurately learn diverse dynamic characteristics spanning different timescales.

Building upon this localized precision, through the hierarchical mechanistic fusion architecture, the framework integrates these local NSSEs into a globally interaction-consistent system model, accurately describing the cross-system interactions among device states across the network nodes. As demonstrated in Fig. 5, this local precision translates seamlessly into highly accurate global trajectory predictions. Quantitative analysis confirms that the step-wise mean absolute error (MAE) remains at a consistently low level without diverging over time. This validates the framework's capability to reconstruct the true underlying dynamic manifold directly from measurement data, realizing high-fidelity prediction and wide-range generalization under complex coupled interactions.

#### 2) Perception and Reconstruction of Explicit State-Space via Hierarchical Analytical Fusion

Due to the physically consistent neural state-space structure design, the HNSSE framework possesses inherent interpretability. Rather than acting as a mere black-box curve fitter, the HNSSE quantitatively captures the underlying physical interaction relationships between state variables. While the model is trained as a dynamic learner to accurately predict state derivatives, the specialized NSSE architecture (as defined in Section III-A) simultaneously compels it to learn the system's underlying dynamic structure. By applying the denormalization process (detailed in Section IV-D) to the outputs of the trained parameter-generating networks, the physical state-space matrices can be recovered for any given

TABLE I
RECONSTRUCTION ACCURACY OF HIERARCHICAL STATE-SPACE MATRICES

| **Element (Component)** | **Learned Value** | **Original Value** | **Element (System)** | **Learned Value** | **Original Value** |
|---|---|---|---|---|---|
| $\partial(\Delta\dot{\delta}_1)/\partial(\Delta\delta_1)$ | -0.0001 | 0.0000 | $\partial(\Delta\dot{V}_{m5})/\partial(\Delta V_{m5})$ | −50.0135 | −50.0000 |
| $\partial(\Delta\dot{\delta}_1)/\partial(\Delta\omega_1)$ | 314.1816 | 314.1593 | $\partial(\Delta\dot{V}_{r1,5})/\partial(\Delta V_{r1,5})$ | −19.9856 | −20.0000 |
| $\partial(\Delta\dot{\omega}_1)/\partial(\Delta\delta_1)$ | -0.2961 | -0.2967 | $\partial(\Delta\dot{V}_{r1,5})/\partial(\Delta V_{f5})$ | −59.9911 | −60.0000 |
| $\partial(\Delta\dot{\omega}_1)/\partial(\Delta\omega_1)$ | -0.5125 | -0.5128 | $\partial(\Delta\dot{V}_{r2,5})/\partial(\Delta V_{r2,5})$ | −1.0001 | −1.0000 |
| System Eigenvalues | 0.0000 | 0.0000 | System Eigenvalues | −0.3716 ±0.5025j | −0.3729 ±0.5008j |
| | −49.9762 | −49.9674 | | −0.2445 ±4.7946j | −0.2445 ±4.7949j |

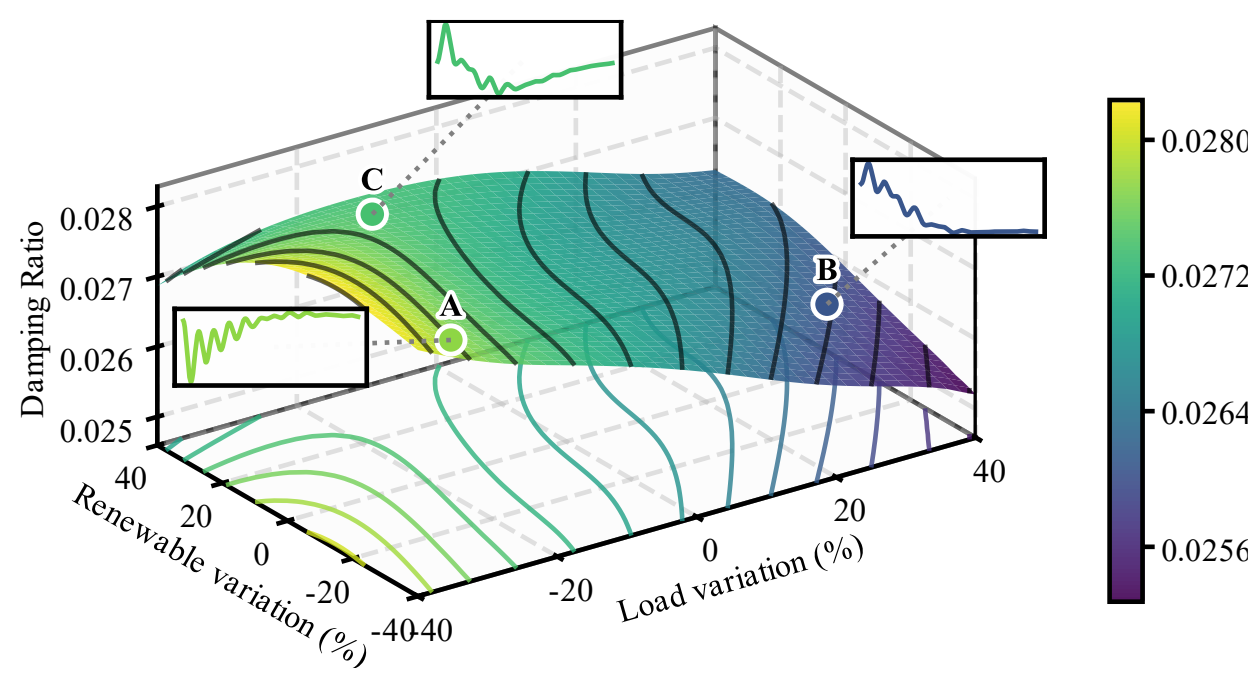

**Fig. 8.** Continuous evolutionary manifold of the damping ratio for the critical oscillation mode under varying renewable energy and load levels.

operating point. This interpretability is assessed at both the component and system levels.

TABLE I provides a representative validation of the structural interpretability at both the local and global levels by comparing representative recovered matrix elements and eigenvalues against the ground truth. At the component level, the learned partial derivatives accurately map the state interactions and closely match the true physical parameters. Furthermore, the precise reconstruction of multi-timescale dynamic properties, ranging from low-frequency electromechanical modes to highly damped fast-response control modes, demonstrates that the framework successfully learns the intrinsic dynamic mechanisms of each respective device. Building upon this local precision, the global system state matrix is constructed by analytically fusing the learned component and network models. The results exhibit a high degree of spectral matching across the entire system manifold, accurately identifying both the monotonic decay modes and the complex electromechanical oscillatory pairs. This precise global alignment confirms that the proposed method effectively captures the complex cross-network couplings among all nodes, achieving an interaction-consistent global state-space representation that bridges data-driven global learning with mechanism-based physical consistency.

*C. Analytical Validation of the HNSSE Continuous Nonlinear Evolution Manifold*

**1) Continuous Stability Perception under Operating Condition Shifts**

To verify the framework's capability to capture continuous nonlinear dynamic evolutions, the global state-space matrices are analytically instantiated across a continuous operational space (spanning a ±40% variation in renewable energy and load levels). By extracting continuous underlying correlations from discrete time-domain measurements, HNSSE successfully constructs a complete evolutionary manifold of system stability. As illustrated in Fig. 8, the damping ratio degrades steadily with increasing load, whereas its response to renewable penetration exhibits a distinct "rise-then-fall" trajectory. This non-monotonic behavior is mechanically driven by the continuous reconfiguration of the synchronizing torque network: moderate renewable integration initially alleviates relative rotor angle stress among dominant synchronous machines, slightly improving damping. Conversely, deeper penetration gradually suppresses conventional generation and amplifies the equivalent synchronizing stiffness, driving the oscillation frequency up faster than the real eigenvalue shift and ultimately causing the overall damping ratio to decline.

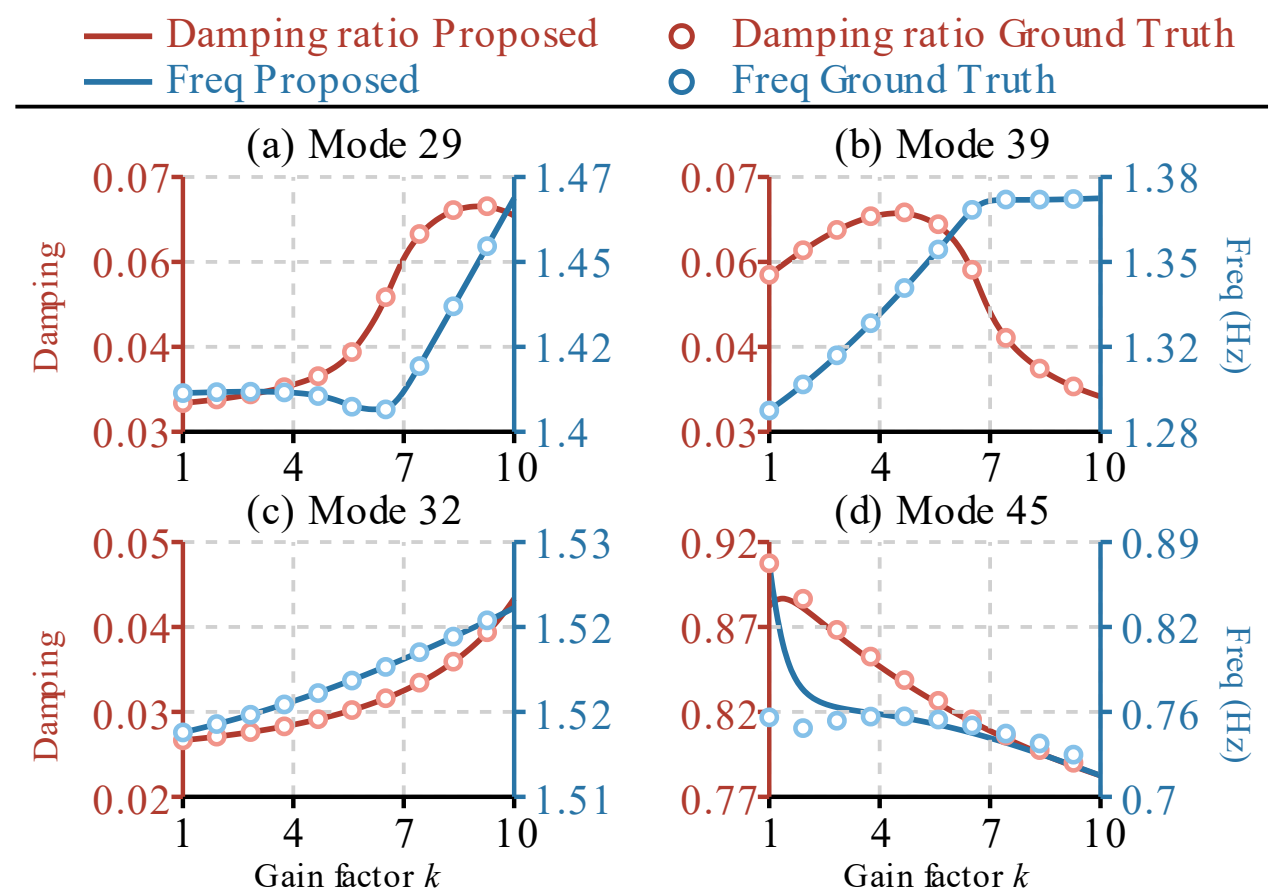

**Fig. 7.** Migration trajectories of damping ratio and frequency for key system modes under varying PSS gain factors.

These reconstructed trends strictly align with both ground-truth samples and theoretical mechanisms. By accurately recovering the continuous dynamic manifold from isolated historical data, the proposed framework explicitly bridges discrete measurements with cross-condition dynamic trends. Practically, this enables the quantitative online perception of continuous stability margins based on real-time operating data, providing reliable analytical support for proactive oscillation risk identification under highly complex grid conditions.

**2) Eigenvalue Migration Tracking for Control Parameter Evaluation**

In addition to capturing dynamic variations caused by shifts in macroscopic operating states, the framework can also effectively track the impact of component-level control parameter settings on global system dynamics. To demonstrate this capability, the PSS gain factor of Generator 5 is continuously varied from 1 to 10. Fig. 7 presents the tracked migration trajectories of both the key electromechanical oscillation modes and the high-frequency control mode.

The results indicate that the neural-predicted root loci tightly align with the ground truth, effectively capturing the highly nonlinear relationships between the control parameter and the system eigenvalues. For the electromechanical oscillation modes (e.g., Modes 29, 32, and 39), while increasing the gain generally enhances the damping, the framework accurately models its distinct nonlinear impacts across different modes. For instance, as the gain factor increases, the damping improvement for certain modes gradually weakens, whereas for Mode 39, it experiences a clear degradation after a peak. Furthermore, the analytical global matrix allows the framework to accurately perceive the reduction in the control stability margin caused by the increased gain. As illustrated in Fig. 7(d) (Mode 45), while a larger PSS gain improves the damping of oscillation modes, the model simultaneously detects a distinct rightward migration of the control mode eigenvalues. Although a slight frequency deviation exists for this high-frequency control mode—primarily due to its marginal contribution to the macroscopic trajectory error—the predicted variation trend remains completely consistent with the ground truth.

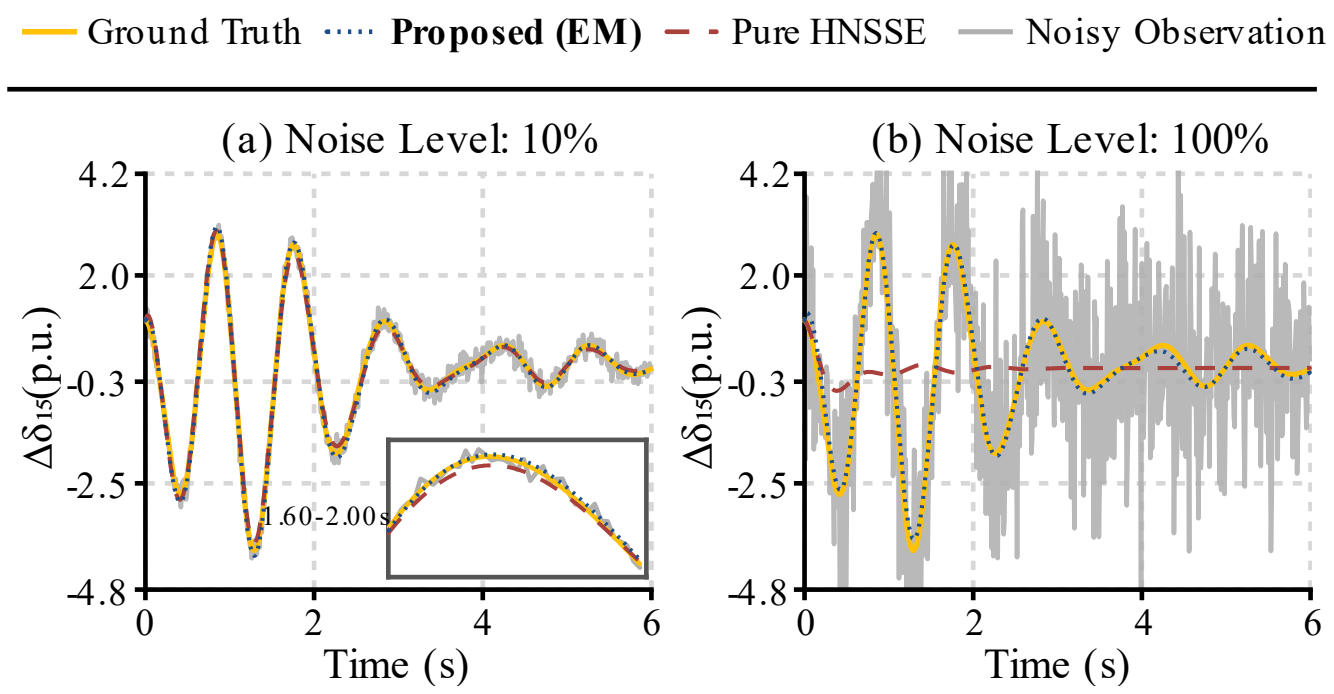

**Fig. 9.** Comparison of trajectory reconstruction between the Pure HNSSE and the proposed EM-refined framework under varying noise levels.

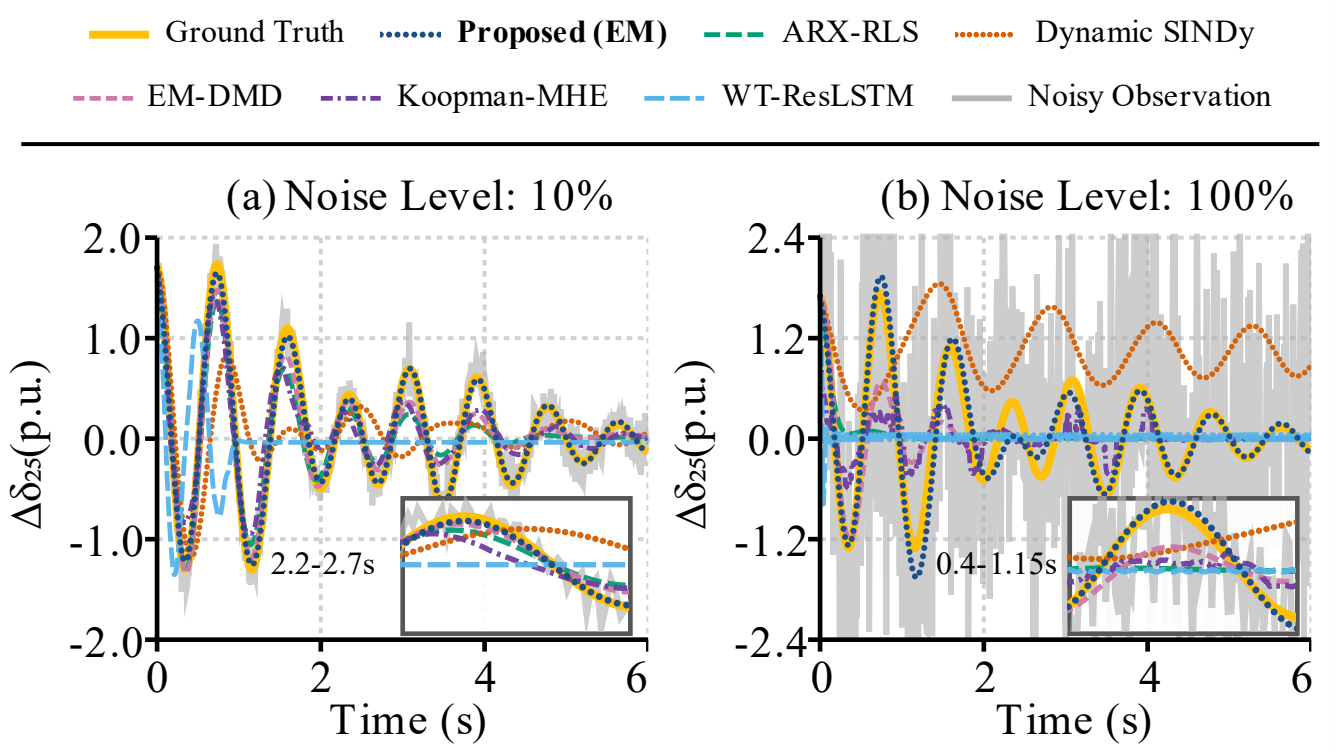

**Fig. 10.** Trajectory tracking performance comparison between the proposed framework and advanced baseline methods under stochastic measurement noise.

By establishing an interaction-consistent coupling relationship between device and system through the analytical fusion mechanism, the proposed framework accurately captures the specific impacts of local device parameters on both the global system oscillation modes and the local control modes. This quantitative tracking capability provides reliable analytical support for the systematic tuning and margin assessment of control parameters in large-scale systems.

### *D. Robustness Evaluation under Stochastic Measurement Noise*

To evaluate the framework's resilience against practical measurement uncertainties, Gaussian white noise was introduced into the training dataset. Performance is benchmarked at noise levels of 10% and 100%, relative to the standard deviation of the normalized state fluctuations.

**1) Efficacy of EM Refinement across Varying Noise Levels**

To evaluate the internal efficacy of the EM refinement mechanism, the proposed framework is compared against the "Pure HNSSE" (without the self-bootstrapping loop), as visualized in Fig. 9. Inherently, the Pure HNSSE establishes a stable trajectory baseline under moderate interference (e.g., 10% noise). This intrinsic resilience stems from a structural synergy: probabilistic slicing captures multi-scale dependencies, the ODE solver provides natural low-pass filtering, explicit physical constraints restrict model complexity, and the neural backbone leverages wide-area samples to smooth stochastic variations. However, under severe 100% noise, the Pure HNSSE trajectory exhibits significant amplitude attenuation (Fig. 9(b)), as the network inadvertently over-smooths heavy fluctuations to minimize prediction loss, suppressing the true transient energy.

Building upon this robust yet attenuated foundation, the Self-Bootstrapping EM strategy perfects the reconstruction. By employing the Pure HNSSE output as a high-quality prior, the framework utilizes the resulting denoised trajectories to iteratively refine the model parameters. Through this cyclic interaction, the mechanism achieves accurate noise separation and high-precision dynamic learning, successfully recovering the full dynamic energy and accurately tracking genuine system transients even under the extreme 100% noise condition.

**2) Benchmark Analysis against Advanced Paradigms**

The proposed framework is benchmarked against five advanced paradigms: Expectation-Maximization Dynamic Mode Decomposition (EM-DMD) [26], Auto-Regressive with Exogenous input (ARX) model with Recursive Least Squares (ARX-RLS) [27], Koopman with Moving Horizon Estimation (Koopman-MHE) [28], Dynamic SINDy for non-stationary sparse ODE discovery (Dynamic SINDy) [29], and Wavelet Transform and Residual Long Short-Term Memory (WT-ResLSTM) [30]. As shown in Fig. 10, the baseline methods experience significant performance degradation under noisy environments, whereas the proposed framework consistently maintains a high degree of alignment with the ground-truth trajectories.

The comparative analysis reveals the distinct structural limitations of existing methods. While EM-DMD is effective at low noise levels, it fails to secure an accurate initial iteration point under severe interference. This validates HNSSE's superiority in utilizing its base dynamic representation as a structurally consistent prior. Furthermore, online recursive approaches like ARX-RLS are highly susceptible to "structural forgetting"; corrupted streaming samples bias the identified parameters, leading to over-smoothed transients. HNSSE circumvents this by explicitly modeling the continuous dynamic mapping across the entire wide-area sample space.

Finally, Koopman-MHE and Dynamic SINDy inherently struggle to distinguish severe stochastic noise from genuine system dynamics. Meanwhile, WT-ResLSTM lacks explicit state-space interpretability and tends to over-smooth physically meaningful oscillations, leading to flattened amplitudes and distorted phases. Conversely, the HNSSE framework imposes explicit state-space structural constraints and hierarchical analytical fusion strictly conforming to the physical network topology. This unique hybrid design inherently regularizes the learning process and integrates multi-dimensional nodal data, ensuring robust, physically consistent trajectory predictions under highly uncertain operational environments.

## VI. Conclusion

This paper proposes the HNSSE framework, which reconciles the analytical transparency of classical physics-based models with the continuous nonlinear generalization of AI. First, the framework employs the NSSE to capture component manifolds through neural regularization that deeply integrates power system mechanisms. Subsequently, it analytically fuses these local dynamics into an interaction-consistent system-level NODE to circumvent the curse of dimensionality. Furthermore, a robust training strategy incorporating probabilistic slicing, multi-stage curriculum learning, and self-bootstrapping refinement ensures high-fidelity reconstruction against noise and multi-scale dynamics, as validated on the real-world Guangdong Power Grid. Future work will extend this foundational platform toward theoretical advancements, such as the deep integration of data and mechanism and global non-linear representation. Additionally, the framework will be applied to downstream applications, including online stability margin assessment and model-based controller design, to support the advanced analysis and control of modern power systems.